\documentclass[aps,prl,twocolumn,showpacs]{revtex4}
\usepackage{amsbsy,latexsym}
\usepackage{amsfonts}
\usepackage[mathscr]{eucal}

\newcommand{\tr}{{\rm tr}\,}
\newcommand{\smfrac}[2]{{\scriptstyle {#1 \over #2}}}

\begin{document}
\title{Optimal strategies for sending information through a
quantum channel}
\author{E.~Bagan, M.~Baig, A.~Brey and R.~Mu\~{n}oz-Tapia}
\affiliation{Grup de F\'{\i}sica Te\`{o}rica \& IFAE,
Facultat de Ci\`{e}ncies, Edifici Cn, Universitat Aut\`{o}noma de
Barcelona, 08193 Bellaterra (Barcelona) Spain}
\author{R.~Tarrach}
\affiliation{Departament d'Estructura i Constituents de la
Mat\`{e}ria, Universitat de Barcelona, Diagonal 647, 08028
Barcelona, Spain}
\date{\today}

\begin{abstract}
Quantum states can be used to encode the information contained in
a direction, i.e., in a unit vector. We present the  best
encoding procedure when the quantum state is made up of $N$ spins
(qubits).  We find that the quality of this optimal procedure,
which we quantify in terms of the fidelity, depends solely on the
dimension of the encoding space. We also investigate the use of
spatial rotations on a quantum state, which provide a natural and
less demanding encoding.  In  this case we prove that the
fidelity is directly related to the largest zeros of the Legendre
and Jacobi polynomials. We also discuss our results in terms of
the information gain.
\end{abstract}


\maketitle

One of the problems which is helping us to deepen our
understanding of quantum information theory is that of sending
information through a quantum channel. Suppose Alice wants to
send to Bob the information contained in an arbitrary direction,
i.e., in a unit vector $\vec{n}$, which she encodes  in a quantum
state. This state is sent to Bob, who performs a quantum
measurement to retrieve the information stored in the state.
Given the characteristics of the information source, and of the
quantum channel, there must exist an optimal encoding-decoding
procedure which maximizes the knowledge Bob can acquire about
$\vec{n}$.

The aim of this letter is to present such optimal
codifications for an isotropic distribution.
After considering in detail the lowest dimensions $d=2$, $3$ and
$4$, we will find the corresponding best procedure for the general
case. Such codification, although mathematically very simple, is
rather difficult to implement physically.
Therefore, we also consider a more
natural strategy: that in which Alice only performs
physical rotations to her code state \cite{MP,DBE,GP,Ma}. In
this case we obtain the optimal strategy for any number of
spins.

Consider the simplest possible quantum channel, of dimension
$d=2$, which can be interpreted as a spin-$1/2$ particle. The
optimal encoding-decoding procedure is the obvious one \cite{MP}:
$\vec{n}$ is encoded in the state of spin pointing into $\vec{n},$
$\vec{\sigma}\cdot \vec{n}| \vec{n}\rangle =|\vec{n}\rangle $,
and the decoding is performed by a standard von Neumann spin
measurement in an arbitrary direction,
$\vec{\sigma}\cdot\vec{m}$. From this measurement, Bob obtains
two possible outcomes, $\pm 1$, to which he associates the guesses
$|\pm \vec{m} \rangle$. We use the fidelity,
$( 1\pm \vec{n}\cdot \vec{m})/2$, as a figure of merit for Bob's
guesses. One
could have taken another measure for quantifying the merit of the
guess, such as the gain of information (or mutual information),
but it complicates substantially the mathematics. Furthermore,
previous work \cite{TV} seems to indicate that the optimal
strategy is insensitive to the use of any of the two figures of
merit. Nevertheless, we present also some comments and results
concerning the information gain at the end of this letter. Now,
let us write the average fidelity (for simplicity, we will
loosely refer to it simply as fidelity)
for $d=2$ as~\cite{MP}
\begin{eqnarray}
  F^{(2)}&&=\int {\rm d}\vec{n}\left[ \frac{1+\vec{n}\cdot
  \vec{m}}{2} \left| \left\langle
  \vec{n}|\vec{m}\right\rangle \right| ^{2} \right. \nonumber \\
  &&+\left. \frac{
  1-\vec{n}\cdot \vec{m}}{2}\left| \left\langle
  \vec{n}|-\vec{m}\right\rangle \right| ^{2}\right]
  =\frac{2}{3},
  \label{f2}
\end{eqnarray}
where we have used $\left| \left\langle
\vec{n}|\vec{m}\right\rangle \right| ^{2}= (1+\vec{n}\cdot
\vec{m})/2$. Notice also that the source, being isotropic, is
characterized by the density matrix, $\rho ^{(2)}=\int {\rm
d}\vec{n}\left| \vec{n}\right\rangle \left\langle \vec{n}\right|
=I^{(2)}/2$, where $I^{(d)}$ is the identity in $d$ dimensions, and,
hence, has maximal von Neumann entropy, $S(\rho
^{(2)})=-\tr \rho ^{(2)}\log _{2}\rho ^{(2)}=1.$ This is likely
to be a feature of optimal encoding, since the Holevo bound
\cite{H1}, which sets an upper limit on the amount of information
accessible to Bob, is precisely, $S(\rho ^{(2)})$ for pure state
encoding (recently, it has also been proven that the bound is
asymptotically achievable \cite{H2}). It is convenient, for what
follows, to trade the von Neumann measurement for a continuous
(i.e. with infinitely many outcomes) positive operator valued
measurement (POVM),
\begin{equation}\label{povm1}
\left| \vec{m}\right\rangle \left\langle \vec{m}\right| +\left|
-\vec{m} \right\rangle \left\langle -\vec{m}\right| =2\int {\rm
d}\vec{m}\left| \vec{m} \right\rangle \left\langle \vec{m}\right|
=I^{(2)},
\end{equation}
which also leads to a maximal fidelity. Notice that this decoding
measurement projects on precisely the same states, and with the
same relative weights, as those used for encoding the direction.
We also recall that for any optimal measurement it is always
possible to design a continuous POVM that is also optimal
\cite{H3}. Therefore, only these need to be considered
to find maximal fidelities, although finite measurements leading
to the same fidelity can be found~\cite{DBE}.

Consider now $d=3$, or a spin-1 particle. It is no longer  obvious
how to encode $\vec{n}$ in an optimal way, since pure states are
now characterized by four parameters, while $\vec{n}$ depends only on
two. (Recall that the code state, $\left| A
(\vec{n})\right\rangle $, can be taken to be pure, as if it
were a mixed state, one could always replace it with the
pure state component which is optimal with respect to the
POVM.) In order to determine $\left| A
(\vec{n})\right\rangle $ we will make the following natural
assumptions: {\it (a)~the optimal code state $\left| A
(\vec{n})\right\rangle $ is an eigenstate of an operator which
can be interpreted as a spin pointing into the direction
$\vec{n}$,
\begin{equation}\label{assumption-1}
\ \vec{S}\cdot \vec{n}\left| A (\vec{n})\right\rangle
=S_{n}\left| A (\vec{n})\right\rangle ,\ \ \ S_{n}\geq 0,\
\end{equation}
where (b)~states corresponding to different directions are
related by the ``generalized rotations" generated by $\vec{S}$}
(these are genuine spatial rotations only if $\vec S$ is the total
spin of the system). Using this, one can easily solve $d=3$, as
there is only one choice for $\vec{S}$: the spin-1 operators; and
only one for $S_{n}$: $S_{n}=1$ (since $S_{n}=0$ is not
one-to-one). As in the case $d=2$, the source is described by a
maximal von Neumann entropy density matrix, $\rho
^{(3)}=I^{(3)}/3$ , $S(\rho ^{(3)})=\log _{2}3$. Recall, however,
that von Neumann spin measurements are no longer optimal, for it
is  known that in this case optimal measurements must contain at
least four projectors~\cite{LPT}. In fact, no optimal
measurement, except for $d=2$, can be of von Neumann
type~\cite{LPT}. It is easy to verify that a continuous POVM,
projecting on precisely the code states, $ 3\int {\rm
d}\vec{m}\left|1,1\vec{m}\right\rangle \left\langle
  1,1\vec{m}\right| =I^{(3)}$, is optimal,
where (and hereafter) we use the notation $\vec
S{}^2|S,S_{n} \vec n\rangle=S(S+1)|S,S_{n} \vec n\rangle$, $
\vec{S}\cdot
  \vec{n}|S,S_{n}\vec{n}\rangle =S_{n}|S,S_{n}\vec{n}
  \rangle$. One finds that the maximal fidelity is
\begin{equation}\label{f3}
  F^{(3)}=3\int {\rm d}\vec{n}\int {\rm d}\vec{m}\frac{1+\vec{n}\cdot
  \vec{m}}{2}\left|
   \langle 1,1\vec{n}| 1,1\vec{m}\rangle \right| ^{2}
   =\frac{3}{4}.
\end{equation}

The problem becomes more complex for $d=4$. There are now two
different interpretations of such a Hilbert space: that of a single
spin-$3/2$ particle or that of two spin-$1/2$ particles. Consider first
the spin-$3/2$ particle interpretation. From~(\ref{assumption-1}),
we see that
either $S_{n}=3/2$\ or $S_{n}=1/2$. The case $S_{n}= 3/2$
parallels qualitatively that just outlined for $d=3$, and gives
for the corresponding optimal measurements,
$F^{(4)}(S=S_{n}=3/2)=4/5$. The choice $S_{n}=1/2$ leads to a lower
fidelity, in spite of the fact that the two encodings have
maximal entropy sources, as for both $\rho ^{(4)}=I^{(4)}/4$, \
$S(\rho ^{(4)})=2$. This can be understood by noticing the
following results,
\begin{eqnarray}
    \left| \left\langle \smfrac{3}{2},\smfrac{3}{2}
    \vec{n}\mid
    \smfrac{3}{2},\smfrac{3}{2}\vec{m} \right\rangle \right| ^{2}&=&\left(
    \frac{1+\vec{n}\cdot \vec{m}}{2} \right) ^{3}, \\
    \text{ \ \ }\left| \left\langle \smfrac{3}{2},\smfrac{1}{2}\vec{n}\mid
    \smfrac{3}{2},\smfrac{1}{2}\vec{m}\right\rangle \right|
^{2}&=&\frac{\left(
    1+\vec{n} \cdot \vec{m}\right) \left( 1-3\vec{n}\cdot
    \vec{m}\right) ^{2}}{ 8}\text{\ .}
\end{eqnarray}
Thus, for $S_{n}=3/2$, the more differs $\vec n$ from~$\vec m$, the
less $|\smfrac{3
}{2},\smfrac{3}{2} \vec{n}\rangle$ and $|\smfrac{3
}{2},\smfrac{3}{2} \vec{m}\rangle$ overlap, i.e., $\left| \left\langle
\smfrac{3}{2},\smfrac{3}{2}\vec{n}\mid \smfrac{3
}{2},\smfrac{3}{2} \vec{m}\right\rangle \right| ^{2}$ is a
monotonous function of $ {(1+\vec{n}\cdot \vec{m})/2}$, while
this is not the case for $ S_{n}=1/2$. This is a particular
instance of a general feature that emerges from our analysis: {\em
(c)~the overlap of the optimal code states corresponding to different
directions, $\left| \left\langle A (\vec{n})|A (\vec{m}
)\right\rangle \right| ^{2}$, should be a monotonous function of
${(1+ \vec{n}\cdot \vec{m})/2}$, ranging from 0 to 1}. The lack of
this feature enables us to discard the choice
$S_{n}=1/2$ without further ado, as we have also verified by explicit
computation.

Let us now go to the two spin-$1/2$ particle interpretation of
$d=4$. Somewhat surprisingly, there are now two possible spin
operators. The first is the obvious total spin operator,
$\vec{S}=\left( \vec{\sigma} \otimes I+I\otimes
\vec{\sigma}\right)/2$, which leads to the Clebsch-Gordan
decomposition {\boldmath{$1 \oplus 0$}}.
In this case, there are two choices consistent with~(\ref{assumption-1}):
\begin{eqnarray}
\label{alice-1}
\left| A (\vec{n})\right\rangle _{1} &=&\left| \vec{n}%
\right\rangle \left| \vec{n}\right\rangle =\left|
1,1\vec n\right\rangle
, \\
\label{alice-2}
 \left| A (\vec{n})\right\rangle _{0} &=&\cos \alpha
\left| 1,0\vec n\right\rangle +\sin \alpha \;e^{i\beta }\left|
0,0\right\rangle ,
\end{eqnarray}
where $\alpha $ and $\beta $ are $\vec{n}$-independent, as
follows from our assumption~{\it (b)}. The
$S_{n}=1$ case reduces to the $d=3$ one, and gives, of course, the
same maximal fidelity~(\ref{f3}). For~$S_{n}=0$, the
overlapping condition~{\em (c)} implies $\cos \alpha =\sin
\alpha =1/{\sqrt{2}}$, which we have explicitly checked to
be indeed the optimal code for the two spin-$1/2$ particle
interpretation. Notice, however, that the density matrix
describing the source no longer has maximal entropy, since
now $S(\rho ^{(3+ 1)})=1+(1/2)\log _{2}3<2$. This implies that
the optimal decoding POVM cannot project on the very same code
states, since the corresponding set of projectors has to be
a resolution of the identity.
Indeed the optimal decoding measurement is given by
$
  4\int {\rm d}\vec{m}\left| B (\vec{m})\right\rangle \left\langle B (
  \vec{m})\right| =I^{(4)}
$,
where
\begin{equation}\label{B(m)}
  \left| B (\vec{m}
  )\right\rangle =\frac{\sqrt{3}}{2}\left| 1,0\vec m\right\rangle +\frac{
   {\rm e}_{{}}^{i\beta }}{2}\left| 0,0\right\rangle ,
\end{equation}
and gives a fidelity
\begin{eqnarray}
F^{(3+1)}&&=4\int {\rm d}\vec{n}\frac{1+\vec{n}\cdot
  \vec{z}}{2}\left|
   \langle A(\vec{n})| B(\vec{z})\rangle \right| ^{2}\nonumber\\
   &&=\frac{3+\sqrt{3}}{6},
   \label{fidelity-3+1}
\end{eqnarray}
where $\vec{z}$ is the unit vector pointing in the $z$ direction
and rotational invariance enabled us to integrate $\vec m$ trivially.
Notice that ${\rm e}^{i\beta }=\pm 1$ corresponds to the code states
chosen by Gisin and Popescu \cite{GP} that led them to the
conclusion that antiparallel spins encode information about
$\vec{n}$ more efficiently than parallel spins. Our result
reproduces theirs, which was later proven to be optimal
\cite{Ma}. Note, however, that the fidelity (\ref{fidelity-3+1})
is lower than $4/5$, the spin-$3/2$ particle interpretation result.

Before discussing our results, let us dispose of the other spin
operators, which are in fact a one-parameter family,
$S_{i}=(\cos ^{2}\eta\, \sigma _{i}\otimes I+\sin ^{2}\eta\,
I\otimes \sigma _{i}+ \sin \eta \cos \eta \, \sum_{j,k}\epsilon
_{ijk}\sigma_j \otimes \sigma_{k})/2$. They generate the
{\boldmath{$1/ 2 \oplus 1/ 2$}} representation, and one can easily check
that $F^{(2+2)}=F^{(2)}$. It is thus of no interest.

We can draw the following conclusion  from our analysis of $d
=4$: the optimal encoding is given by the spin-$3/2$
interpretation, i.e., by the only encoding which satisfies
Eq.~(\ref{assumption-1}), the overlapping condition~{\em (c)},
and corresponds to a maximal entropy
source. This is, after all, what one would have expected.
This result can be generalized to an arbitrary dimension: the single
spin-$(d-1)/2$ interpretation of a $d$-dimensional Hilbert space gives
the optimal encoding with maximal fidelity
\begin{equation}
    F^{(d)}={d\over d+1}.
    \label{d-dim}
\end{equation}
 If $d=2^{N}$, one can,
of course, perform this optimal encoding with $N$ spin-$1/2$
particles (qubits).

Let us now illustrate this for the simple case of two qubits: the
operators $S_{i}$ corresponding to the spin-$3/2 $
interpretation, can be written as \cite{Pe} $S_x
=(\sqrt{3}/{2})\, I\otimes \sigma_x + (\sigma_x\otimes \sigma_x
+\sigma_y\otimes \sigma_y)/2$, $S_y = (\sqrt{3}/{2}) \,I\otimes
\sigma_y + (\sigma_y\otimes \sigma_x -\sigma_x\otimes
\sigma_y)/2$, and $S_z=(1/2)\, I\otimes \sigma_z+\sigma_z \otimes
I$. These operators fulfill the $SU(2)$ algebra, $[S_i,S_j]=i
\epsilon_{ijk}S_k$, but they are not the components of a vector
under spatial rotations, generated by the total spin of the two
particles (we have already found that the only vector
representations are {\boldmath{$1 \oplus 0$}} and {\boldmath{$1/
2 \oplus 1/ 2$}}). The unitary transformations generated by these
operators are non-local and  difficult to implement physically.
Furthermore, they can change the entanglement of the states. For
instance, the product state $\mid
\smfrac{1}{2},\smfrac{1}{2}\vec{z} \rangle \otimes
    \mid
\smfrac{1}{2},\smfrac{1}{2}\vec{z}\rangle$, which is an
eigenvector of $S_z$, becomes entangled under the transformation
${\rm e}^{i \theta S_y}$ for $\theta=\pi/2$, but remains a
product state for $\theta=\pi$.
The optimal decoding  can be achieved by a continuous POVM, but
there are finite POVM's that are optimal too~\cite{DBE,LPT}. For
example, from Ref.~\cite{LPT} one can read off that the minimal
optimal POVM corresponds to six equally weighted projectors
associated to the six unit vectors pointing at the vertices of a
regular octahedron.

The merit of the procedure just outlined is, obviously, that the
maximum possible value of the fidelity is attained. However, the
encoding process, involving complicated unitary operations, looks
exceedingly demanding. It is therefore important 
to examine a
less contrived method in which Alice can only perform spatial
rotations on an initial code state: she may, e.g., rotate the device
that produces her initial states. This is, actually, the
approach followed in~\cite{MP,DBE} for parallel spin code states,
where the maximum fidelity in terms of the number of spins
was found to be $F=(N+1)/(N+2)$, and in~\cite{GP,Ma} for two
antiparallel spins. In fact, for two spins we have already found
that the family of states~(\ref{alice-2}) with $\alpha=\pi/4$
(to which the two antiparallel spin state of~\cite{GP,Ma} belongs),
is indeed the best Alice can use if she is only allowed to perform
space rotations.
We will now generalize this physically more feasible
strategy to any number of spins and calculate its maximal fidelity.

Let us sketch the main steps of the calculation (a more detailed
discussion will be presented elsewhere \cite{BBBMT}). First,
one considers, as usual, continuous POVMs for decoding.
Second, note
that according to the Clebsch-Gordan decomposition,
any state of $N$ spin-$1/2$
particles  can be written as a combination of states
$|S,S_{n}\vec{n}\rangle$,
$0\le S\le N/2$,
belonging to the irreducible representation {\boldmath{$\bf S$}} (here $S$,
$S_{n}$ obviously refer to the total spin operator), where
{\boldmath{$\bf S$}} usually appears more than once for $S<N/2$.
 Third, one notices that these repeated
representations do not add any further knowledge about~$\vec n$,
hence, the Hilbert space, ${\mathscr H}$, of the
code states can be chosen to be
\begin{equation}
{\mathscr H}=   \mbox{\boldmath{$\displaystyle{{\bf N}\over 2}$}}
        \mbox{\boldmath{$\oplus$}}
        \mbox{\boldmath{$\displaystyle \left({{\bf N}\over 2}-1\right)$}}
        \mbox{\boldmath{$\oplus$}}
        \mbox{\boldmath{$\displaystyle \left({{\bf N}\over
        2}-2\right)$}}+\cdots.
    \label{CGtrunc}
\end{equation}
States living in
more than one equivalent representation can also be used, but
this just complicates the computation and leads to the same
maximal fidelity.
According to Eqs.~(\ref{assumption-1}) and~(\ref{CGtrunc}), the
optimal code state can be written as $|A(\vec n)\rangle
=\sum_{S=S_{n}}^{N/2} A_S|S,S_{n}\vec n\rangle$, where
$\sum_{S=S_{n}}^{N/2}
|A_S|^2=1$. One must choose the minimal possible value of $S_{n}$, that is,
$S_{n}=0$ if $N$ is even, and $S_{n}=1/2$ if $N$ is odd, since
these choices  use the largest available dimension of the code state
space~(\ref{CGtrunc})~\cite{BBBMT}. The explicit calculation of the
fidelity function corresponding to the optimal POVM, for which
$|B(\vec m)\rangle$ is
a straightforward generalization of~(\ref{B(m)}), leads to
\begin{equation}\label{fidelity-general}
F=\frac{1}{2}+\frac{1}{2} \mathsf{A}^t \mathsf{M}  \mathsf{A},
\end{equation}
where  $\mathsf{M}$ is a matrix of tridiagonal form
\begin{equation}\label{matrix}
\mathsf{M}=\pmatrix{d_{l}&c_{l-1}&
         &        &       \cr
                    c_{l-1}&\ddots&\ddots
    &\phantom{\ddots} \raisebox{2.0ex}[1.5ex][0ex]{\LARGE 0}\hspace{-.5cm}       &       \cr
                         &\ddots&d_{3}    &c_{2}  &       \cr
                         & \phantom{\ddots}    &c_{2}&d_{2}&c_{1}      \cr
\hspace{.5cm} \raisebox{2.0ex}[1.5ex][0ex]{\LARGE 0}\hspace{-.5cm}
&&\phantom{\ddots}&c_{1}&d_{1}}
\end{equation}
that can be
chosen to be real. Here
\begin{equation}
l=N/2+1-S_n ,
\label{n-N}
\end{equation}
and $\mathsf{A}^t=(|A_{N/2}|,|A_{N/2-1}|,|A_{N/2-2}|,\dots)$,
where $\mathsf{A}^t$ is the transpose of $\mathsf A$. If $N$ is
even, the coefficients of $\mathsf M$ are $d_k=0$,
$c_{k}=k/{\sqrt{4 k^2-1}}$, otherwise, if $N$ is odd, $d_k=1/(4
k^2-1)$, $c_{k}={\sqrt{k(k+1)}}/({2k+1})$. The largest eigenvalue,
$x_{l}$, of $\mathsf M$ determines the maximal fidelity through
the relation $F=(1+x_l)/2$. To find $x_{l}$, we set up a
recursion relation for  the characteristic polynomial of $\mathsf
M$:
\begin{equation}\label{ch-p}
Q_{l}(x)=(d_l-x) Q_{l-1}(x) -c_{l-1}^2 Q_{l-2}(x).
\end{equation}
We are now at the end of the calculation, as the solutions of~(\ref{ch-p})
are just proportional to the Legendre polynomials,
$P_l(x)$, if $N$ is even, and to the Jacobi
polynomials~\cite{AS}, $P^{0,1}_l(x)$, if $N$ is odd. The
eigenvalue $x_l$ is precisely the largest zero of
the corresponding polynomial.

The values of the maximal fidelity for $N$ up to seven are collected in
Table~I.
\begin{table}\label{table-I}
\begin{center}
\begin{tabular}{c|ccccccc}
  \toprule
  $N$ & 1 & 2 & 3 & 4 & 5 & 6 & 7 \\
  \colrule
  $F$ & $\frac{2}{3}$ & $\frac{3+\sqrt{3}}{6}$ & $\frac{6+\sqrt{6}}{10}$ &
  $\frac{5+\sqrt{15}}{10}$
  & $.9114$ & $.9306$ & $.9429\phantom{\Big|}$\\
  \botrule
\end{tabular}
\end{center}
\caption{Maximum fidelities in terms of the number of spins for
space rotations}
\end{table}
\indent Notice that the  optimal encoding for three spins gives
$F=(6+\sqrt{6})/10 \sim 0.845 $, which is  a better result than
the optimal value for 4 parallel spins ($F=5/6\sim .833 $
\cite{MP}). In fact, it can be shown that our maximal fidelity
approaches unity quadratically in the number of spins:
\begin{equation}\label{assymptotic-1}
  F\sim 1-\frac{\xi^2}{N^2},
\end{equation}
where $\xi\sim 2.4$ is the first zero of the Bessel function
$J_0(x)$, while for $N$ parallel spins the fidelity approaches unity only
linearly: $F\sim 1-1/{N}$.
At this point, we feel compelled to go back to~(\ref{d-dim}) and
point out that for the optimal encoding, based on generalized
rotations, the fidelity tends exponentially to unity:
$F\sim1-2^{-N}$.

Up to now  we have restricted ourselves to finding  optimal
strategies using the fidelity to quantify the quality of the
encodings. We
would like to conclude by making a few comments on their
quantum information gain. We, therefore, work out this quantity
for the optimal strategies that led to~(\ref{d-dim}).

For a  continuous POVM the symmetry of the problem enables us to
simplify the computation, as only the contribution of a single
projector is needed (say the one in the $\vec z$ direction). After
canceling the divergent terms associated to the continuous
distribution of the code states $|A(\vec n)\rangle$,
the average information gain
is just~\cite{BBBMT}
\begin{eqnarray}
\label{information-gain-1}
I_{\rm av}=\int {\rm d}&&\vec n \left( d\;|\langle A(\vec n)|B(\vec
z)\rangle|^2\right)\nonumber\\
&&\times\log
_{2}\left( d\;|\langle A(\vec n)|B(\vec
z)\rangle|^2\right),
\end{eqnarray}
where
$|B(\vec z)\rangle=\left| S,S\, \vec{z}\right\rangle$, and $d$,
the
dimension of the code state space,
is related to $S$ by $d=2S+1$. In terms of $d$  one obtains
$I_{\rm av}=\log _{2}d-(1-1/d)\log_{2}{\rm e}$,
a result also found in \cite{TV}. In terms of $N$, it reads
\begin{equation}\label{information-gain-2}
I_{\rm av}=N-(1-2^{-N})\log_{2}{\rm e}.
\end{equation}
This is just the number of qubits transmitted in the process,
minus a term that asymptotically goes to a constant.

Finally, it is interesting  to study the information gain using
the simpler, but not truly optimal, encoding. For $N=2$, the best code state
according to the fidelity is given by
(\ref{alice-2}) with $\alpha/\pi=1/4$ (maximal fidelity
and information gain are
both independent of $\beta$). The information gain is
$I_{\rm av}=0.8664$, less than that obtained
applying the optimal encoding for which~(\ref{information-gain-2})
gives $I_{\rm av}=$ $0.9180$.
Nevertheless, we could ask ourselves if this gain is maximal
for code states of the form~(\ref{alice-2}).
An explicit computation shows that this
is not so, as the maximal gain is $I_{\rm av}=0.8729$ for
$\alpha/\pi=0.2317\not=1/4$. Hence, at least in this
case, states with maximal fidelity and maximal information gain
do not coincide,
they seem to do so only when the truly optimal strategy is considered.

To summarize, we have presented  optimal encoding-decoding
procedures for sending the information contained in an arbitrary
direction faithfully codified in a quantum state.
For restricted encodings, based on space
rotations, the maximal
fidelity is related to the largest zeros of the
Legendre or Jacobi polynomials. Although this encoding does not
make full use of the quantum channel capacity, our results show a
significant improvement
over previous strategies based on parallel spin encoding.

We thank S.~Popescu, A.~Bramon, G.~Vidal and
W.~D\"ur for stimulating discussions, and M.~Lavelle
for reading the manuscript. Financial support from
CICYT contracts AEN98-0431, AEN99-0766, CIRIT contracts
1998SGR-00026, 1998SGR-00051, 1999SGR-00097 and EC contract
IST-1999-11053 is acknowledged.

\end{document}